\documentclass[10pt]{eptcs} 
\usepackage[T1]{fontenc}
\usepackage{tabularx}
\usepackage{graphicx}
\usepackage{times}
\usepackage[latin1]{inputenc}
\usepackage{fancyhdr}
\usepackage{hyperref}
\usepackage{xcolor}               
\usepackage[english]{varioref}   
\usepackage{subfigure}
\usepackage{tabularx}
\usepackage{amssymb}
\usepackage{fancyvrb} 
\usepackage{multirow}

\usepackage{pgf}
\usepackage{tikz}
\usepackage[ruled,linesnumbered,lined]{algorithm2e}

\usetikzlibrary{arrows,automata,calc}
\usetikzlibrary{positioning}

\hypersetup{colorlinks=true}

\tikzstyle{loc}=[draw,circle,minimum size=7mm,inner sep=1pt]
\tikzstyle{inloc}=[draw,circle,double,minimum size=7mm,inner sep=1pt]
\tikzstyle{location}=[draw,circle,inner sep=5pt,minimum size=7mm] 
\tikzstyle{tlocation}=[draw,circle,fill=blue!20,inner sep=5pt,minimum size=7mm] 
\tikzstyle{ulocation}=[draw,circle,fill=blue!50,inner sep=5pt,minimum size=7mm] 
\tikzset{font=\sf,>=latex'}

\let\DELTA=\Delta
\def\Delta{\ensuremath{\DELTA}}

\newcommand{\name}{\mathit{SpecName}}

\newcommand{\periodic}{\mathsf{Periodic}}
\newcommand{\aperiodic}{\mathsf{Aperiodic}}
\newcommand{\first}{\mathsf{First}}
\newcommand{\both}{\mathsf{Both}}
\newcommand{\priority}{\mathsf{Priority}}
\newcommand{\memory}{\mathsf{Memory}}
\newcommand{\rendering}{\mathsf{Rendering}}

\newcommand{\compactlist}{\setlength{\itemsep}{1pt}}

\newcommand{\ie}                 {i.e., }
\newcommand{\eg}                 {e.g., }

\newcommand{\drop}[1]            {}

\newcommand{\mrl}{\textsc{MIRELA}}

\newcommand{\tast}{{\sc TAST}} 
\newcommand{\uppaal}{{\sc UPPAAL}} 
\newcommand{\prism}{{\sc PRISM}} 
\newcommand{\onlyS}{\cal {O}} 
\newcommand{\noSDU}{\cal {N}}
\newcommand{\SDU}{\cal {W}}
 
\newcommand{\tastsem}{\mathit{tasts}} 
\newcommand{\prismsem}{\mathit{prism}}  
\newcommand{\spec}{{\cal S}}	

\newif\ifcomment
\commenttrue 
\definecolor{gris}{gray}{0.3}

\newcommand{\mycomment}[3]{\ifcomment
 {\small \null\-
  {\color{#1}{\textbf{#2:}} \color{#1} {#3}} }
 \fi
}
\newcommand{\EndSy}{\hfill\protect\makebox[1.0em][c]{
\protect\setlength{\unitlength}{0.2em}
\protect\begin{picture}(3,3)(0,0)
        \begin{thinlines}
\protect\put(0,0){\line(1,0){3}}
\protect\put(0,0){\line(0,1){3}}
\protect\put(0,3){\line(1,0){3}}
\protect\put(3,0){\line(0,1){3}}
\protect\put(0,0){\line(1,1){3}}
\protect\put(0,3){\line(1,-1){3}}
        \end{thinlines}
\protect\end{picture}
\protect\setlength{\unitlength}{1mm}
}}
\newcommand{\BX}[1]{{\unskip\nobreak\hfil\penalty50
                    \hskip2em\hbox{}\hfil
\EndSy\/ {{\rm #1}}
                    \parfillskip=0pt \finalhyphendemerits=0 \par
                   }}
\newcommand{\PROP}[2]{\goodbreak\begin{proposition}
                      \label{#1}{\sc #2}

                     }
\newcommand{\ENDPROP}{
\end{proposition}}
\newcommand{\ENXPROP}[1]{\BX{\ref{#1}}
\end{proposition}
                       }
\newcommand{\PROOF}{\goodbreak
{\bf Proof:}
                 }
\newcommand{\ENDPROOF}[1]{\BX{\ref{#1}}
                       }

\newcommand{\HK}[1]{\mycomment{red!80!black}{HK}{#1}}

\newtheorem{example}{Example}{\itshape}{\rmfamily}
\newtheorem{proposition}{Proposition}{\bfseries}{\itshape}

\title{Indefinite waitings in \mrl{} systems}

\author{
Johan Arcile\qquad\qquad  Jean-Yves Didier\qquad\qquad  Hanna Klaudel
  \institute{Laboratoire IBISC, Universit\'e d'Evry-Val d'Essonne, France}
  \email{johan.arcile@ens.univ-evry.fr\qquad\qquad  \{jean-yves.didier,hanna.klaudel\}@ibisc.fr}
\and
  Raymond Devillers
   \institute{D\'epartement d'Informatique,\\ Universit\'e Libre de Bruxelles, Belgium}
   \email{rdevil@ulb.ac.be}
\and
  Artur Rataj
  \institute{Institute of Theoretical and Applied Computer Science,\\ Gliwice, Poland}
  \email{arturrataj@gmail.com}
}
\def\titlerunning{Indefinite waitings}
\def\authorrunning{J. Arcile, R. Devillers, J-Y. Didier, , H. Klaudel \& A. Rataj}

\begin{document}
\maketitle
\setcounter{totalnumber}{6}

\begin{abstract} 
\drop{The \mrl{} framework aims at easing the development of 
concurrent applications, 
from a high level specification to a low level implementation. 
}
\mrl{} is a high-level language and a rapid prototyping framework 
dedicated to systems 
where virtual and digital objects coexist in the same environment and interact
in real time.
Its semantics is given in the form of networks of timed automata, 
which can be checked using symbolic methods. 
This paper shows how to detect various kinds of
indefinite waitings in the components of such systems. 
The method is experimented using the \prism{} model checker. 
\end{abstract}

%


\section{Introduction}
The aim of this paper is to provide a formal method support 
for the development of concurrent applications, which consist of components, 
which mutually interact in a way, that should meet certain real--time
constraints, like a reaction time within a given time period.
\mrl{} (for MIxed REality LAnguage~\cite{mir08,sand13,mir13,issre14}) 
was initially meant to be used for developing mixed reality (MR)
~\cite{JY} applications, which acquire data from sensors 
(like cameras, microphones, GPS, haptic arms\ldots),
and then distribute it through a shared memory, read by
rendering devices, which present the results 
in a way a human can interpret (using senses like sight, hearing, touch; 
by highlighting images on a screen, projecting virtual 
images, mixing virtual and real images, moving robot arms\ldots).

One of the ideas behind \mrl{} is to translate a model into an equivalent form
understandable by a model checker, as opposed to performing state space
exploration like \eg JPF does \cite{visser2003modelchecking}.
It allows the
model to be represented not by a transition matrix, possibly very lengthy
and still partial, but by a terse specification in the checker's native
input language. The checker may then easily apply symbolic data structures
like MTBDDs \cite{kwiatkowska2004probabilistic}, which may in turn allow for
a substantial reduction of the space explosion problem, inherent to an explicit
transition matrix.

In order to cope with time constraints when developing MR applications, 
practitioners rely mostly on fast response and high performance hardware, 
even if this contradicts other issues, like power saving (hence autonomy) 
and cost (hence mass production), 
and does not ascertain that critical constraints will always be respected. 
Modelling the application before testing it on the actual hardware 
and validating it by applying formal method techniques to prove its robustness, 
was the main motivation for the development of the \mrl{} framework.
It aims at supporting the development process of MR set-ups, which
are generally prone to various issues related to time and known to be difficult to 
control and to adjust. 
Most mixed reality frameworks, like those cited in 
\cite{bauer01,reitmayr01,haller03,piekarski03,hughes2005mixed,endres05} 
do not concentrate on the validation of the developed applications. 
Some of them emphasise the use of formal descriptions of components 
in order to enforce a modular decomposition \cite{sandor2001,latoschik2002}, 
and ease future extensions \cite{navarre2005}
or substitutions of one module by another 
\cite{figueroa2004,figueroa2008}.
Such frameworks do not deal with software failure issues related to time.
On the contrary, the main focus of the \mrl{} framework is the formal analysis 
of software failure issues related to time, together with timing performance analyses and
the development of automatic tools.

The \mrl{} framework~\cite{sand13} proposes a methodology that consists of three
phases.
In the first phase, a formal specification of the system in the form of a network of timed automata~\cite{alur:90} is built. 
It may be obtained by a translation from a high level description made of connected components
\cite{mir08,mir13}, 
and represents an ideal world.
The second phase concerns the analysis of the system: it essentially consists 
in analysing through model-checking a set of desired  properties considered important, 
either the absence of bad behaviours or the satisfaction of timing constraints.
In the third phase, such a checked specification is used to produce an implementation skeleton, 
in the form of a looping controller parametrised with a sampling period and possibly executing 
several actions in the same period, aiming at preserving those properties \cite{sand13}.
We revisit here essentially the second phase of the methodology of the \mrl{} framework. 
Since the high level specifications of \mrl{} are close to a subclass of \uppaal~\cite{uppaal4.1}
systems, it  was originally considered to use the \uppaal~model-checker to analyse the 
properties of a \mrl{} system \cite{issre14}.
However, a serious problem was faced when trying to detect deadlocks limited to some components, since the 
\uppaal~query language does not allow nested path quantifiers. 
A proposed solution was then to use instead the \prism{} tool \cite{kwiatkowska2004probabilistic}
and analyse if and how it may be to detect bad behaviours of a tentative \mrl{} system.

The paper is organised as follows. 
First, we shortly recall the specification language of \mrl{} and its semantics in 
terms of a network of \tast{} automata (a subclass of timed automata of \uppaal). 
Next, various kinds of bad behaviours are defined.
Then, Section \ref{sec:prism_semantics} explains how \prism{} may be used to model \mrl{} systems, 
and the next one analyses how to verify the system against the indefinite waitings phenomena.
This leads to define a procedure to analyse a \mrl{} system, which is illustrated on a well chosen example.
Finally we summarise the outcome of this contribution, and comment some future works.

\subsection{\mrl{} syntax and intuitive semantics} 
\label{stx.sect}

\label{stx.sect.syntax}

A \mrl{} specification~\cite{sand13} (see an example in Figure \ref{fig:example1_tast}, top left), 
is defined as a list of component's declarations of the form:
\[ \begin{array}{l}
\name{:}  	\hspace{.5cm} id = \mathit{Comp}{\rightarrow}\mathit{TList}; \ldots ; 
	~id = \mathit{Comp}{\rightarrow}\mathit{TList}.
\end{array}
\]
Each component's declaration  $\mathit{Comp}{\rightarrow}\mathit{TList}$
defines a component $\mathit{Comp}$ and its target list of components $\mathit{TList}$,
which is an optional (comma separated) list of identifiers  
indicating to which (target) components information is sent, 
and in which order.
Each component also indicates from which (source) components data are expected.
A target $t$ of a component $c$ must have $c$ as a source, but 
it is not 
required that a source $s$ of a component $c$ has $c$ as an explicit target: 
missing targets will be implicitly added at the end of the target list, 
in the order of their occurrence in the specification list. 
We assume that all the sources of a component are different, and that all the targets of 
a component are also different\footnote{The target list 
		is allowed to be empty; this defines in general a degenerate specification, 
		which may be interesting for technical and practical reasons.}. 
A component $\mathit{Comp}$ is either 
a sensor $\mathit{Sensor}$, 
a processing unit $\mathit{PUnit}$, 
a shared memory unit $\mathit{MUnit}$ or 
a rendering loop $\mathit{RLoop}$, and is specified following the syntax:
\[
\begin{array}{lll}
\mathit{Sensor} &:: =& \periodic(\mathit{min\_start},\mathit{max\_start})[\mathit{min} {,} \mathit{max}] \mid  \aperiodic(\mathit{min\_event}) \\

\mathit{PUnit} &:: = & \first(SList) \mid  
	  \both(id,id)[\mathit{min}{,}\mathit{max}] \mid  
           \priority(id[\mathit{min}{,}\mathit{max}],id[\mathit{min}{,}\mathit{max}]) \\
 
\mathit{MUnit} &:: =& \memory(\mathit{SList}) \\

\mathit{RLoop} &:: = & \rendering(\mathit{min\_rg}, \mathit{max\_rg})(id[\mathit{min}{,}\mathit{max}]),

\end{array}
\]
where
$\mathit{SList}$ is a non empty list of (comma separated) 
source identifiers of the form 
$\mathit{id}[\mathit{min}{,}\mathit{max}]$, indicating that the processing time 
of data coming from source $id$ takes between  $\mathit{min}$ and $\mathit{max}$ 
time units. 

There are two kinds of sensors: 
\begin{itemize}\compactlist
	\item $\periodic$ ones (\eg cameras) 
	that need some time for being started (at least
	$\mathit{min\_start}$ and at most $\mathit{max\_start}$ time units), and then 
	capture data periodically, taking between $\mathit{min}$ and at most 
	$\mathit{max}$ 
	time units for that, and
	\item $\aperiodic$ ones  (\eg haptic arms or graphical user interfaces) 
	that collect data when an event occurs, 
	the parameter $\mathit{min\_event}$ indicating the minimal delay between taking two successive events into account.
\end{itemize}
 
Processing units 
process data coming from possibly several different sources of data. 
They may be combined (in a hierarchy but also in loops) to get more inputs and outputs. 
Hence the sources are either sensors or processing units, and targets are either  
memories or processing units.
There are the following categories of processing units: 
\begin{itemize}\compactlist
\item	$\first$: may have one or more inputs (sources) and starts processing when data are received from one of them; the order is irrelevant; 
if $\mathit{SList}$ contains only one element, $\first$ is considered as a unary
processing unit;

\item	$\both$: has exactly two inputs and starts processing when both input data are received, the processing time being between $\mathit{min}$ and $\mathit{max}$; 
\item	$\priority$: has two inputs (master and slave) and starts processing when the master input is ready, possibly using the slave input if it is available before the master one; 
the duration of processing is in the first time interval $[\mathit{min}{,}\mathit{max}]$ if the master input is alone available, and in  the second  time interval $[\mathit{min}{,}\mathit{max}]$ if both the slave and the master inputs are captured; in figures, the slave input is indicated by a dashed arrow.
\end{itemize}

A memory access is performed by a rendering loop, a sensor or a processing unit 
by locking the memory before executing the corresponding task (reading or writing) followed by an unlocking of the memory. 
A rendering component accesses the memory at a flexible period between 
$\mathit{min\_rg}$ and  $\mathit{max\_rg}$ time units, and the processing of data 
has a duration in the interval $[\mathit{min}{,}\mathit{max}]$.

\begin{example}{}
\label{ex1}
Let us consider the example corresponding to the \mrl{} system specified in Figure \ref{fig:example1_tast} 
(top left), with three periodic sensors feeding two $\first$ processing units, 
and a $\both$ one fed by the last sensor and the last $\first$, and with 
a single rendering unit with its associated memory.
The corresponding flow of information is illustrated in Figure \ref{fig:example1_tast} (top right).\BX{\ref{ex1}}
\end{example}

\begin{example} {}
\label{ex2}
This is a variant of Example \ref{ex1} where 
$R  = \mathsf{Rendering}(50,75)(M[25,50])$ 
is replaced by $R  = \mathsf{Rendering}(75,100)(M[25,50])$,
\ie the rendering time is longer. 
The flow of information in this model is the same as in Example \ref{ex1}. \BX{\ref{ex2}}
\end{example}

\begin{figure}[htb!]
\centering 
\begin{tabular}{cc}

\begin{minipage}{0.55\textwidth}
$\begin{array}{rcl}
Ex_1{:} && \\
  S1  &=& \mathsf{Periodic}(50,75)[75,100]; \\
  S2 &=& \mathsf{Periodic}(200,300)[350,400] {\rightarrow}(F2,F1); \\
  S3  &=& \mathsf{Periodic}(200,300)[350,400] {\rightarrow}(F2,B); \\
  F1  &=& \mathsf{First}(S1,S2[50,75]); \\
  F2  &=& \mathsf{First}(S2,S3[75,100]); \\
  B  &=& \mathsf{Both}(S3,F2)[25,50]; \\ 
  M  &=& \mathsf{Memory}(F1[25,50],B[25,50]); \\
  R  &=& \mathsf{Rendering}(50,75)(M[25,50]). \\ 
\end{array}
$
\end{minipage}

&

\begin{minipage}{0.5\textwidth}
\begin{tikzpicture}[xscale=.7, yscale=.9] 

\node at (6,3) {sensors};
\draw[blue!10,fill] (-1,0.5) rectangle (8,2.5); 
\node at (6,1.5) {processing units};
\node at (6,0) {shared memories};
\draw[blue!10,fill] (-1,-1.5) rectangle (8,-.5); 
\node at (6,-1) {rendering loops};

\draw (0,3) node[rectangle,draw,thick,rounded corners] (s1) {$S1$};
\draw (2,3) node[rectangle,draw,thick,rounded corners] (s2) {$S2$};
\draw (4,3) node[rectangle,draw,thick,rounded corners] (s3) {$S3$};
\draw (1,2) node[rectangle,draw,thick,rounded corners] (f1) {$F1$};
\draw (3,2) node[rectangle,draw,thick,rounded corners] (f2) {$F2$};
\draw (4,1) node[rectangle,draw,thick,rounded corners] (b) {$B$};
\draw (2,0) node[rectangle,draw,thick,rounded corners] (m) {$M$};
\draw (2,-1) node[rectangle,draw,thick,rounded corners] (r) {$R$};

\draw[thick,->] (s1) -- (f1) ;
\draw[thick,->] (s2) -- (f1) ;
\draw[thick,->] (s2) -- (f2) ;
\draw[thick,->] (s3) -- (f2) ;
\draw[thick,->] (s3) -- (b) ;
\draw[thick,->] (f1) -- (m) ;
\draw[thick,->] (f2) -- (b) ;
\draw[thick,->] (b) -- (m) ;
\draw[thick,->] (m) -- (r) ;
\end{tikzpicture}

\end{minipage}
\end{tabular}\\[2mm]

\begin{tabular}{l}
\tikzstyle{location}=[draw,circle,inner sep=5pt] 
\begin{tikzpicture}[>=latex',xscale=0.5, yscale=.6,every node/.style={scale=0.7}]
\node[tlocation,double] at (-4,0) (I) {$s_0$}; 
\node[tlocation]  (s1) {$s_1$};
\node[location] at (4,0) (s2) {$s_2$};
\node[above] at (I.north) {$x < 75$}; 
\node[above] at (s1.north) {$x< 100$};
\draw[->] (I) -- (s1) node[midway,below] {$x := 0$}node[midway,above] {$x\ge 50$}  ;
\draw[->] (s1) -- (s2) node[midway,above] {$x \ge 75$} ;
\draw[->,rounded corners] (s2) -- ++(0,-1.5) -| (s1) node[above,near start] {$k_{S1-F1} !$} node[below,near start] {$x:= 0$};
\node at (-3,-1.5) {\begin{tabular}{l} Sensor ${S1}$ \end{tabular}} ;
\end{tikzpicture} 
\hspace{3mm}
\tikzstyle{location}=[draw,circle,inner sep=5pt] 
\begin{tikzpicture}[>=latex',xscale=0.6, yscale=.6,every node/.style={scale=0.7}]
\node[tlocation,double] at (-4,0) (I) {$s_0$}; 
\node[tlocation]  (s1) {$s_1$};
\node[location] at (4,0) (s2) {$s_2$};
\node[location] at (2,-1.5) (s3) {$s_3$};
\node[above] at (I.north) {$x < 300$}; 
\node[above] at (s1.north) {$x < 400$};
\draw[->] (I) -- (s1) node[midway,below] {$x:= 0$} node[midway,above] {$x \ge 200$};
\draw[->] (s1) -- (s2) node[midway,above] {$x \ge 350$};
\draw[->,rounded corners] (s2) -- ++(0,-1.5) -- (s3)  node[above,midway] {$k_{S2-F2} !$} ;
\draw[->,rounded corners] (s3) -- ++(-2,0) node[midway,below] {$x:= 0$}  node[above,midway] {$k_{S2-F1} !$} -- (s1)   ;
\node at (-3,-1.5) {\begin{tabular}{l} Sensor ${S2}$ \end{tabular}} ; 
\end{tikzpicture} \\

\tikzstyle{location}=[draw,circle,inner sep=5pt] 
\begin{tikzpicture}[>=latex',xscale=0.6, yscale=.6,every node/.style={scale=0.7}]
\node[tlocation,double] at (-4,0) (I) {$s_0$}; 
\node[tlocation]  (s1) {$s_1$};
\node[location] at (4,0) (s2) {$s_2$};
\node[location] at (2,-1.5) (s3) {$s_3$};
\node[above] at (I.north) {$x < 300$}; 
\node[above] at (s1.north) {$x < 400$};
\draw[->] (I) -- (s1) node[midway,below] {$x:= 0$} node[midway,above] {$x \ge 200$};
\draw[->] (s1) -- (s2) node[midway,above] {$x \ge 350$};
\draw[->,rounded corners] (s2) -- ++(0,-1.5) -- (s3)  node[above,midway] {$k_{S3-F2} !$} ;
\draw[->,rounded corners] (s3) -- ++(-2,0) node[midway,below] {$x:= 0$}  node[above,midway] {$k_{S3-B} !$} -- (s1)   ;
\node at (-3,-1.5) {\begin{tabular}{l} Sensor ${S3}$ \end{tabular}} ;
\end{tikzpicture} 
\hspace{3mm}
\tikzstyle{location}=[draw,circle,inner sep=5pt] 
\begin{tikzpicture}[>=latex',xscale=0.6, yscale=.6,every node/.style={scale=0.7}]
\node[location,double] at (0,0) (s0) {$s_0$};
\node[tlocation] at (4,0) (s1) {$s_1$};
\node[location] at (8,-2) (s2) {$s_2$};
\node[tlocation] at (4,-2) (s3) {$s_3$};
\node[location] at (0,-2) (s4) {$s_4$};
\node[below] at (s1.south) {$x < 75$};
\node[below] at (s3.south) {$x < 50$};
\draw[->,rounded corners] (s0) -- ++(0,1.5) -| (s1) node[above,near start] {$k_{S1-F1}?$} node[below,near start] {$x := 0$};
\draw[->] (s0) -- (s1) node[above,midway] {$k_{S2-F1}?$}
  node[below,midway,sloped] {$x := 0$};
\draw[->,rounded corners] (s1)  -|   (s2) node[above,near start] {$x\ge 50$};
\draw[->] (s2) -- (s3) node[above,midway] {$lock!$} node[below,midway]{$x:=0$} ;
\draw[->] (s3) -- (s4) node[above,midway] {$x\ge 25$};
\draw[->] (s4) -- (s0) node[left,midway] {$unlock!$};
\node at (0,-3) { $\first ~F1$} ;
\end{tikzpicture} \\[.3cm]
%
\tikzstyle{location}=[draw,circle,inner sep=5pt] 
\begin{tikzpicture}[>=latex',xscale=0.6, yscale=.6,every node/.style={scale=0.7}]
\node[location,double] at (0,0) (s0) {$s_0$};
\node[tlocation] at (4,0) (s1) {$s_1$};
\node[location] at (8,0) (s2) {$s_2$};
\node[below] at (s1.south) {$x < 50$};
\draw[->,rounded corners] (s0) -- ++(0,1.5) -| (s1) node[above,near start] {$k_{S2-F2}?$} node[below,near start] {$x := 0$};
\draw[->] (s0) -- (s1) node[above,midway] {$k_{S3-F2}?$}
  node[below,midway,sloped] {$x := 0$};
\draw[->,rounded corners] (s1) -- (s2) node[above,midway] {$x\ge 25$};
\draw[->,rounded corners] (s2) -- ++(0,-2) -| (s0) node[above,near start] {$k_{F2-B}!$};
\node at (1,-3) { $\first ~F2$} ;
\end{tikzpicture}
\begin{tikzpicture}[>=latex',xscale=0.6, yscale=.6,every node/.style={scale=0.7}] 
\node[location,double] at (0,-1) (W) {$s_0$};
\node[tlocation] at (8,-1) (P) {$s_3$};
\node[location] at (4,.5) (n) {$s_1$};
\node[location] at (4,-1) (nn) {$s_2$};
\node[location] at (8,-3) (m) {$s_4$};
\node[tlocation] at (4,-3) (n1) {$s_5$};
\node[location] at (0,-3) (o) {$s_6$}; 
\node[right] at (P.east) {$x < 50$};
\node[below] at (n1.south) {$x < 50$};
\draw[->,rounded corners] (W) |- (n) node[above,near end] {$k_{F2-B}?$}; 
\draw[->,rounded corners] (n) -| (P) node[below,near start] {$x:= 0$}node[above,near start] {$k_{S3-B}?$};
\draw[->] (W) -- (nn) node[above,midway] {$k_{S3-B}?$};
\draw[->] (nn) -- (P) node[above,midway] {$k_{F2-B}?$}node[below,midway] {$x := 0$};
\draw[->] (P) -- (m) node[right,midway] {$x\ge 25$};
\draw[->] (m) -- (n1) node[above,midway] {$lock!$}node[below,midway] {$x:=0$};
\draw[->] (n1) -- (o) node[above,midway] {$x\ge 25$};
\draw[->] (o) -- (W) node[left,midway] {$unlock!$};
\node at (0,-4) {$\both~B$} ;
\end{tikzpicture}\\[0.2cm]
\hspace{1cm}
\begin{tikzpicture}[>=latex',xscale=0.6, yscale=.6,every node/.style={scale=0.7}] 
\node[location,double] (W) {$s_0$};
\node[location] at (3,0) (A) {$s_1$};
\draw[->,rounded corners] (W) -- ++(0,1.5) -| (A) node[above,near start] {$lock?$};
\draw[->,rounded corners] (A) -- (W) node[above,midway] {$unlock?$};
\node at (1.5,-1) {Memory $M$} ;
\end{tikzpicture}
\hspace{1cm}
\begin{tikzpicture}[>=latex',xscale=0.5, yscale=.6,every node/.style={scale=0.7}] 
\node[location,double] (W) {$s_0$};
\node[tlocation] at (4,0) (M) {$s_1$};
\node[location] at (8,0) (A) {$s_2$};
\node[tlocation] at (4,-2) (R) {$s_3$};
\node[above] at (M.north) {$x < 50$};
\node[above] at (R.north) {$x < 75$};
\draw[->] (W) --  (M) node[above,midway] {$lock!$} node[below,midway] {$x:=0$};
\draw[->] (M) --  (A) node[above,midway] {$x\ge 25$};
\draw[->,rounded corners] (A) |-  (R) node[above,near end] {$unlock!$} node[below,near end] {$x:=0$};
\draw[->,rounded corners] (R)  -| (W) node[above,near start] {$x\ge 50$};
\node at (4,-3) {Rendering $R$} ;
\end{tikzpicture}

\end{tabular}\\

\caption{\label{fig:example1_tast} Specification, abstract scheme and {\tast} representation for
Example \ref{ex1}. For Example \ref{ex2}, the {\tast} representation is as for Example \ref{ex1}
except for the invariant of location $s_3$ in $\rendering$ $R$ (which becomes $x<100$) 
and the guard of its out-going arc (which becomes $x\geq 75$). 
}
\end{figure}

\label{sec:tasts}

Originally, the semantics of a \mrl{} specification has been defined and implemented in \uppaal{} \cite{uppaal4.1}
as a set of timed automata \cite{alur:90,alur:91,alur:94,Waez2011} with urgent binary synchronisations, 
meaning that when a synchronisation is possible, time may not progress. 
More precisely, we used a subclass called Timed Automata with Synchronised Tasks (\tast{}s) in order to cope with implementability issues 
(see  \cite{sand13} for more details). 

Syntactically, 
a \tast{} is an annotated directed (and connected) graph, with an initial node, 
provided with a finite set of non-negative real variables called \emph{clocks} (\eg $x$), initially set to $0$, 
increasing with time and reset ($x:=0$) when needed. 
Clocks are not allowed to be shared between automata. 
The nodes (called \emph{locations}) are annotated with \emph{invariants} 
(predicates allowing to enter or stay in a location, typically either empty (meaning true or $x<\infty$) 
or of the form $x < e'$, where $e'$ is a natural number. 
The locations associated with an internal activity (called \emph{activity locations}) are distinguished from the 
locations where one waits for some event or contextual condition (called \emph{wait locations}). 
In figures, locations will be represented by round nodes, the initial one having a double boundary,
and activity locations are indicated by a coloured background. 
The arcs are annotated with \emph{guards} (predicates allowing to perform a move) or \emph{communication actions}, 
and possibly with some clock \emph{resets}. 
For an activity location, all output arcs have a guard of the form $x\geq e$, 
all input arcs reset $x$ and the invariant is either empty or of the form $x < e'$, with $0<e<e'$. 
For a waiting location, all the output arcs  
have a communication action of the form $k!$ (output) or $k?$ (input), allowing to glue together the various automata composing a system, 
since they must occur by input-output pairs. Recall that synchronisations are assumed to be \emph{urgent}, 
which means that they take place
without time progression.
In order to structurally avoid Zeno evolutions 
(i.e., infinite histories taking no time or a finite time), we assume that each loop 
in the graph of the automaton presents (at least) a constraint $x{\geq}e$ in
a guard ($e$ is strictly positive) and a reset of $x$ for some clock $x$,
or contains only input channels ($k?$). 

A \tast~representation of Example~\ref{ex1} is depicted in Figure \ref{fig:example1_tast} (bottom). 
The translation from a \mrl{} specification to a \tast~model (and hence a gateway to the
usage of \uppaal{} or \prism{} for model-checking the system) has been automated 
by developing a compiler using a parametric approach~\cite{arcile2014}. 


\section{Bad behaviours} 
\label{sec:bad}

In \cite{issre14}, we analysed the various kinds of bad behaviours that can occur in a timed system in general (and in a \mrl{} one in particular). 
For instance, one may distinguish:
\begin{itemize}
\item a \emph{complete blocking} occurs if a state is reached where nothing can happen:
 no location change is nor will be allowed (because no arc with a true guard is available 
or the only ones available lead  to locations with a non-valid invariant) and the time is blocked 
(because the invariant of the present location is made false by time passing);
\item a \emph{global deadlock} occurs when only time passing is ever allowed: no 
location change is nor will be possible; 
\item a \emph{strong  (resp. weak) Zeno} situation occurs when infinitely many location changes may be done without time passing (resp. in a finite time delay);  
\item a \emph{local deadlock} occurs if  no location  change is available for some component 
while other components may evolve normally;
\item a \emph{starvation} occurs at some point  if a component may evolve
 but the time before may be infinite, because other components may delay it indefinitely; 
\item  an  \emph{unbounded waiting} occurs if a component  
may eventually evolve but the time before is unbounded, because some activity is unbounded. 
\end{itemize} 
 
We shall denote by \emph{indefinite waiting} those last three situations. 
Note that those situations are not always to be considered as bad: it depends on their semantical interpretation. For instance if a part of a system corresponds to the handling of an error, 
it may be valid that the system stops after the handling, and it is hoped that it is possible to never reach this situation.

Moreover, we have shown \cite{issre14} that, for a \mrl{} system,
\begin{itemize}
\item no (strong or weak) Zeno situation may happen;
\item  a component may only deadlock in a waiting location;
\item  a memory unit may only deadlock if all its users deadlock elsewhere; 
\item  a rendering loop may not deadlock, so that a system with a rendering loop may not present a global deadlock. 
\end{itemize} 

As a consequence,  a global deadlock may not occur in a complete system, 
\ie having at least one memory unit and an associated rendering loop;
but it can occur in a degenerate (or simplified) system without (memory and) 
rendering loop. 
On the contrary, local deadlocks may occur even in complete systems and may propagate to other components.
%
A component may starve for example if it tries to send information to a memory or to 
a $\first$ component which is continually used by other units, and no fairness strategy is applied.
From these properties it is sometimes possible to reduce the detection of local deadlocks of a 
\mrl{} system to a global deadlock analysis (easy and efficient with \uppaal) of  a reduced 
systems, obtained by dropping the memory units and the rendering loops, and the timing 
constraints as well \cite{issre14}. However, this does not work in all circumstances and
we shall now examine how \prism{} may be used for that purpose.


\section{\prism{} representation of \mrl} 
\label{sec:prism_semantics}

\prism{} \cite{kwiatkowska2004probabilistic} is  a probabilistic model checker intended to analyse a wide variety of
systems, including non-deterministic ones. Hence,
\tast{}s and more generally timed automata are particular cases of models 
\prism{} is able to handle. 
Furthermore, and this is the most interesting feature of \prism{} in what we are concerned here, 
it can use complex (nested) CTL formulas that \uppaal{} cannot.
However \prism{} accepts models that are slightly different from the ones used in \uppaal.
In particular:
\begin{itemize}
\item Communication semantics: in \uppaal, communications are performed through
binary (input/output) synchronisations on some channel $k$.
A synchronisation transition triggers simultaneously exactly one pair of edges $k?$ and $k!$, 
that are available at the same time in two different components. 
\prism{} implements n-ary synchronisations, where an edge labelled $[k]$ may only occur 
in simultaneity with edges labelled $[k]$ in all components where they are present; 
\item Urgent channels: \uppaal{} offers a modelling facility
by allowing to declare some channels as urgent. 
Delays must not occur if a synchronisation transition on an urgent channel is enabled. 
\prism{} does not have such a facility
and thus it should be "emulated" using a specific construct compliant with \prism{} syntax;
\item Discrete clocks: 
the \prism's tool allowing to check CTL formulas is the {\em digital clocks engine}, which uses discrete clocks only (and consequently excludes strict inequalities in the logical formulas).
This modifies the semantics of the systems, but it may be considered that continuous time, as used by \uppaal, is a mathematical artefact and that the true evolutions of digital systems are governed by discrete time devices.

\drop{ 
\item Strict inequalities in location invariants: \prism{} has several model checking engines for
timed automata. However, the one allowing nested CTL formulas relies on digital clocks
\cite{kwiatkowska2006digital} and thus do not allow strict inequalities in location invariants.
}
\end{itemize}


Implementing binary communications in \prism{} is easy 
by demultiplying and renaming channels in such a way that 
a different synchronous channel $[k]$ is attributed to each pair $k?$ and $k!$ of 
communication labels.
In \mrl{} specifications, the only labels we have to worry about  
are the $lock?$ and $unlock?$  labels in each $\memory$ $M$ and the
$lock!$ and $unlock!$ labels in the components that communicate with $M$.

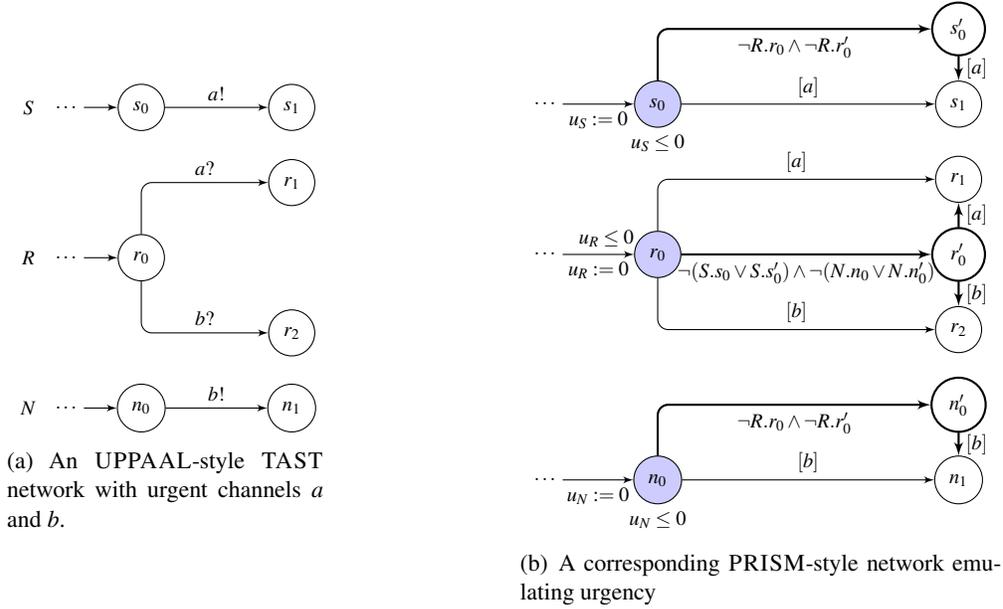
\begin{figure}[tb!]
\begin{center}

\begin{tabular}{cc}
\begin{minipage}{0.4\textwidth}

\subfigure[An \uppaal{}-style \tast{} network with urgent channels $a$ and $b$. 
\label{fig:urge1}]{
\begin{tikzpicture}[>=latex',xscale=1, yscale=1,every node/.style={scale=0.7}] 
\node[location] at (0,5) (s0) {$s_0$};
\node[location] at (2,5) (s1) {$s_1$};
\node at (-1.5,5) (s) {$S$};
\node at (-1,5) (d) {$\cdots$};
\draw[->] (d) -- (s0) ;
\draw[->] (s0) -- (s1) node[above,midway] {$a!$};

\node[location] at (0,3) (r0) {$r_0$};
\node[location] at (2,4) (r1) {$r_1$};
\node[location] at (2,2) (r2) {$r_2$};
\node at (-1.5,3) (r) {$R$};
\node at (-1,3) (dr) {$\cdots$};
\draw[->] (dr) -- (r0) ;
\draw[->,rounded corners] (r0) |- node[above,near end] {$a?$} (r1) ;
\draw[->,rounded corners] (r0) |- node[above,near end] {$b?$} (r2) ;

\node[location] at (0,1) (n0) {$n_0$};
\node[location] at (2,1) (n1) {$n_1$};
\node at (-1.5,1) (n) {$N$};
\node at (-1,1) (dn) {$\cdots$};
\draw[->] (dn) -- (n0) ;
\draw[->] (n0) -- (n1) node[above,midway] {$b!$};
\end{tikzpicture}
}
\end{minipage}
&
\begin{minipage}{0.6\textwidth}

\subfigure[A corresponding \prism-style network emulating urgency \label{fig:urge2}]{
\begin{tikzpicture}[>=latex',xscale=1, yscale=1,every node/.style={scale=0.7}] 
\node[tlocation] at (0,5) (s0) {$s_0$};
\node[below] at (s0.south) {$u_S \leq 0$};
\node[location] at (4,5) (s1) {$s_1$};
\node[location,thick] at (4,6) (s0p) {$s_0'$};
\node at (-1.5,5) (d) {$\cdots$};
\draw[->] (d) -- (s0) node[below,midway] {$u_S:=0$} ;
\draw[->] (s0) -- (s1) node[above,midway] {$[a]$} ;
\draw[->,rounded corners,thick] (s0) |- (s0p) node[below,near end] {$\lnot R.r_0 \land \lnot R.r_0'$};
\draw[->,thick] (s0p) -- (s1) node[right,midway] {$[a]$};

\node[tlocation] at (0,3) (r0) {$r_0$};
\node[left] at (r0.north west) {$u_R \leq 0$};
\node[location] at (4,4) (r1) {$r_1$};
\node[location] at (4,2) (r2) {$r_2$};
\node[location,thick] at (4,3) (r0p) {$r_0'$};
\node at (-1.5,3) (dr) {$\cdots$};
\draw[->] (dr) -- (r0) node[below,midway] {$u_R:=0$} ;
\draw[->, rounded corners] (r0) |- (r1) node[above,near end] {$[a]$}  ;
\draw[->,rounded corners] (r0) |- (r2) node[above,near end] {$[b]$}  ;
\draw[->,thick] (r0) -- (r0p) node[below,midway] {$\lnot (S.s_0 \lor S.s_0') \land \lnot (N.n_0 \lor N.n_0')$};
\draw[->,thick] (r0p) -- (r1) node[right,midway] {$[a]$};
\draw[->,thick] (r0p) -- (r2) node[right,midway] {$[b]$};

\node[tlocation] at (0,0) (n0) {$n_0$};
\node[below] at (n0.south) {$u_N \leq 0$};
\node[location] at (4,0) (n1) {$n_1$};
\node[location,thick] at (4,1) (n0p) {$n_0'$};
\node at (-1.5,0) (dn) {$\cdots$};
\draw[->] (dn) -- (n0) node[below,midway] {$u_N:=0$} ;
\draw[->] (n0) -- (n1) node[above,midway] {$[b] $}  ;
\draw[->,rounded corners,thick] (n0) |- (n0p) node[below,near end] {$\lnot R.r_0 \land \lnot R.r_0'$};
\draw[->,thick] (n0p) -- (n1) node[right,midway] {$[b]$};

\end{tikzpicture}
}
\end{minipage}
\end{tabular}

\end{center}
\caption{\label{urgent.fig} Urgent communications in \prism. Thick locations and arcs are the added ones.} 
\end{figure}

The implementation of urgency is much more intricate, 
especially if the objective is to be transparent for 
the execution and also as much as possible for model-checking performance.
The solution we adopt here consists in the following construction, illustrated in Figure~\ref{urgent.fig}. 
Let ${\cal A}=A_1,\ldots,A_n$ be a network of \tast{} components $A_i$. 
We assume that ${\cal A}$ is already renamed in order to implement binary communications.  
For each $A \in {\cal A}$ we declare an additional clock $x_A$ and
 for each location $loc$ in $A$ with outgoing communication edges  to locations $loc_1$, \ldots, $loc_m$, 
labelled respectively $[k_1],\ldots,[k_m]$:
      
         \begin{itemize}
             \item we add the invariant $x_A \leq 0$ to $loc$ and the reset $x_A:=0$ to all input arcs to $loc$;
             \item we introduce a new location $loc'$ and an edge from $loc$ to $loc'$ 
             with a guard $\lnot (g_{loc}^{k_1}  \lor \ldots \lor g_{loc}^{k_m})$, where
             $g_{loc}^{k_i} \equiv A_j.l_{k_i} \lor
                      A_j.l_{k_i}'$ with $A_j.l_{k_i}$ being the (unique) location with outgoing 
             communication edge labelled $[k_i]$ in some other automaton $A_j$, 
             and $A_j.l_{k_i}'$ is the corresponding added location;
             \item for all $i=1,\ldots, m$, we add an edge from $loc'$ to $loc_i$, labelled $[k_i]$.
         \end{itemize}

\begin{proposition}\label{P1.prop}
The \prism{} system so constructed presents the same behaviour as the original \tast{} one. 
\end{proposition}
\PROOF (sketch) The general idea behind the construction above is the following: 
When the control arrives at location $loc$ in some automaton $A$, 
since invariant $x_A \leq 0$ on $loc$  requires no time progression, two cases are possible: 
\begin{itemize}
\item either at least one synchronisation on some $[k_i]$ is immediately possible and one of them must be performed, 
\item or no synchronisation is possible yet and the control passes to $loc'$ where it may wait as long as one of the synchronisations, 
let us say $[k_i]$, becomes available. 
\end{itemize}
The latter occurs when the control arrives at a location $l$ having an outgoing arc labelled $[k_i]$,
in some automaton $A_j$. 
The synchronisation on $[k_i]$ must be performed without 
time progression due to invariant $x_{A_j} \leq 0$ on $l$. 
\ENDPROOF{P1.prop}

\drop{ 
In order to avoid the problem arising from strict inequalities, we shall in the following assume we only consider specifications with closed intervals, but we shall come back again to this issue at the end of the next section.
}

\section{Detection of indefinite waitings in \mrl{} specifications} 
\label{sec:verif}

In order to detect an indefinite waiting at a location \emph{loc} in a component of the \tast{} 
representation $\tastsem(\spec)$ of a \mrl{} specification $\spec$, we may check the CTL formula 
\[ \phi_{loc} = \mbox{EF }  \mbox{EG } {\emph loc}, \]
which checks if there exist a path leading to a situation (EF) such that from there 
it may happen that the component stays (EG)  in location \emph{loc}. 
Since there is no Zeno situation, this may only correspond to an indefinite waiting.
 If $\phi_{loc}$ is false, then there is neither a starvation nor an unbounded waiting nor a deadlock  in {\emph loc}.
If {\emph loc} is the activity location of an aperiodic sensor, we know that there is an unbounded waiting, 
and it is not necessary to perform the model checking  for that. 
The other interesting cases correspond to waiting locations $w$, from which communications $k!$ or $k?$ only are offered.

If we want to delineate more precisely what happens, we may use a query 
\[ \psi_w = \mbox{EF }  \mbox{AG } w, \]
which checks if there is a situation where 
the considered component reached $w$ but there is no way to get out of it: this thus corresponds to a local deadlock. 
From previous observations, it is not useful to apply it to a $\memory$ component if it has corresponding $\rendering$ loop(s), 
nor to locations waiting for an {\em unlock}, even if there is no $\rendering$. 


If $\phi_w$ is true and $\psi_w$ is false, we know that $w$ corresponds to a starvation or an unbounded waiting. 
But if both are true, it may still happen that, while $w$ corresponds to a local deadlock for some reachable environment, 
it is also possible that for another environment, it corresponds to a starvation or an unbounded waiting. 
This uncertainty may  be solved by checking the formula 
\[ \rho_w = \mbox{EF }  \mbox{EG } (w \land (\mbox{EF } \lnot w)), \]
which checks if we can reach a situation where the considered component is in location $w$,
it is possible to indefinitely stay in $w$ (while other components may progress) but it is also possible to escape from $w$. 
This corresponds to a starvation situation or an unbounded waiting.


As we mentioned in the introduction, \uppaal{} does not support nested CTL queries like
$\phi_w$, $\psi_w$ and $\rho_w$. 
On the contrary, we may check them with \prism{}, on the 
\prism{} representation $\prismsem(\spec)$ of $\spec$. 
Indeed, $\prismsem(\spec)$ only differs from $\tastsem(\spec)$ in that each wait location $w$ in $\tastsem(\spec)$ 
is split in two locations $w$ and $w'$ in $\prismsem(\spec)$, such that it is not possible to stay in $w$; 
if the synchronisation is not performed at $w$ with no time progression, 
$\prismsem(\spec)$ goes to $w'$, where one shall wait as in $w$ in $\tastsem(\spec)$.
The problem thus comes down to check $\phi_{w'}$, $\psi_{w'}$ and $\rho_{w'}$ on $\prismsem(\spec)$. 

In order to distinguish starvation from unbounded waitings 
(i.e., if some wait location incurs starvation only, unbounded waiting only, or both in different environments), 
let us assume the considered specification $\spec$ presents 
$n$ aperiodic sensors (with $n>0$, otherwise there are trivially no unbounded waitings),
and let us denote by $a_1, a_2,\ldots, a_n$ their respective initial locations in $\tastsem(\spec)$. 
 To check a starvation in location $w$, we may use on $\tastsem(\spec)$ 
 the following query formula:
\[ \sigma_w= \mbox{EF }\mbox{EG }(w\land (\mbox{EF }\neg w)\land(\mbox{F }\neg a_1)
\land\cdots\land (\mbox{F }\neg a_n))\]
which means it is possible to stay indefinitely in $w$, but also to escape from it, 
without needing that an aperiodic sensor (or many of them)  indefinitely stays in its activity location. Hence, if true, this means there is a pure starvation in $w$.
To check an unbounded waiting in the same location, one may use on $\tastsem(\spec)$ 
the query:
\[\zeta_w=\mbox{EF }((\mbox{EG }w) \land \mbox{A}((\mbox{G }w)\Rightarrow
(\mbox{FG } a_1)\lor\cdots\lor(\mbox{FG }a_n)))\]
which means it is is possible to stay indefinitely in $w$, but not without being stuck in some $a_i$ at some point.
If this is true, this thus means we have an unbounded wait in $w$.
Unfortunately, those last two formulas belong to CTL*
and presently, when considering non-deterministic properties, 
 \prism{} only supports a fragment of CTL, so that operators G and F must be used 
in alternation with operators A and E. Hence, $\sigma_w$, which contains GF, and $\zeta_w$, 
which contains FG, 
are queries that \prism{} does not support (yet).


Let us note that, if we were to introduce probabilities in the model, it is very likely that starvations  will disappear almost surely (i.e., with probability 1). Indeed, they correspond to the indefinite reproduction of a same kind of finite evolution, and the probability of it is zero, unless that kind of finite evolution has probability 1. Similarly, the probability of unbounded waitings should be zero, like the probability of staying indefinitely in some activity location of an aperiodic sensor (otherwise one could not qualify the waiting as unbounded instead of infinite).

\subsection{Procedure and experimental results} 

From a graph analysis of the specification one may observe that
a location in some component may be concerned by starvation, 
local deadlock or unbounded waiting if it is either a wait location or the initial activity location of an aperiodic sensor. 
Also, among all the wait locations, we can distinguish the following families: 
  \begin{itemize}
    \item the set $\noSDU$ of wait locations that are origins of $\mathit{unlock?}$ or $\mathit{unlock!}$ transitions:
             these are concerned by neither local deadlocks nor unbounded waitings nor starvation;  
             
    \item the set $\onlyS$ of wait locations that are origins of $\mathit{lock!}$ transitions: 
            these cannot be concerned by local deadlocks nor by unbounded waitings,
             but may potentially be concerned by starvation;

    \item the set $\SDU$ of all remaining wait locations.

   \end{itemize}

\begin{proposition} \label{propo-static}
Let $\spec$ be a \mrl{} specification. 
\begin{enumerate}
\item If $\spec$ comprises an aperiodic sensor, it contains by construction at least 
a location concerned by an unbounded waiting (which may propagate to other components). 
On the contrary, if $\spec$ has no aperiodic sensor, no unbounded waiting may occur. 
\item If $\spec$ contains an unbounded waiting in some wait location $w$ in $\tastsem(\spec)$, $\spec$ has aperiodic sensors and $w\in\SDU$.

\item If $\spec$ contains a starvation in some wait location $w$ in $\tastsem(\spec)$, $w\in\onlyS\cup\SDU$.
\item If $\spec$ contains a local deadlock in some wait location $w$ in $\tastsem(\spec)$, $w\in\SDU$.
\item A wait location $w$ in $\tastsem(\spec)$ incurs a local deadlock, a starvation or an unbounded waiting 
iff the same occurs in the corresponding location $w' $ in $\prismsem(\spec)$.
\end{enumerate}
\end{proposition}

\PROOF 
\begin{enumerate}
\item  By definition, an aperiodic sensor contains a location, in which it may be stuck from the very beginning. 
It is also the only way to introduce a location where an unbounded waiting is allowed.
\item See the previous point. 

\item No location $w$ that is the origin of an {\em unlock!} or {\em unlock?} 
may be concerned by a starvation because this would mean that the memory, 
once engaged with a rendering or a processing unit, could be indefinitely waiting.
As renderings and processing units may never be indefinitely waiting between having performed a {\em lock!} on the memory
and the corresponding  {\em unlock!}, $w\not \in \noSDU$.
However, one may  be indefinitely waiting while trying 
to perform a {\em lock!} on a memory or a communication action with a component, 
but only because the memory (in case of a {\em lock!}) or the component is continually working for someone else.
\item See the previous points.
\item The only difference in the \tast{} semantics and the \prism{} one is that each wait location 
$w$ is split into two locations $w$ and $w'$, and it is not possible to stay in $w$: 
if the rendez-vous is not performed at $w$ without any delay, 
\prism{} goes to $w'$, where one shall wait as in $w$ in the \tast{} model.
\end{enumerate}
\ENDPROOF{propo-static}

\drop{
\PROOF 
\begin{enumerate}
\item  An aperiodic sensor may be stuck from the very beginning in its activity location.
\item See the previous point. 
\item Since each memory has a corresponding $\rendering{}$, 
neither the $\rendering{}$
nor the memory may be indefinitely blocked, and the other components may not be indefinitely 
blocked while trying to perform an {\em unlock!}; one may have an indefinite blocking while trying 
to perform a {\em lock!}, but only because the memory is continually working for someone else.
\item See the previous points.
\item The only difference in the \tast{} semantics and the \prism{} one is that each wait location 
$w$ is split into two locations $w$ and $w'$, and it is not possible to stay in $w$: 
if the rendez-vous is not performed at $w$ without any delay, 
\prism{} goes to $w'$, where one shall wait as in $w$ in the \tast{} model.
\end{enumerate}
\ENDPROOF{propo-static}
}

\begin{algorithm}[htb!] 
\KwData{$\SDU$, $\onlyS$ -- sets of wait locations of a  
\mrl{} specification}
\KwResult{Compute, for each wait location, if it is a starvation, an unbounded waiting, a (local) deadlock or a combination of them.}
\BlankLine
\ForEach{$w \in \SDU\cup \onlyS$}{
    Check $\phi_w \leftarrow \mbox{EF EG }w$\;
    \eIf{ $\phi_w = \mbox{false}$}{$w$ is neither a starvation, unbounded waiting nor deadlock;\ }
    {
        \eIf { $w \in \onlyS$}{$w$ is a starvation location}
       {      Check $\psi_w \leftarrow \mbox{EF AG }w$\;
               \eIf{$\psi_w = \mbox{false} $}{
                $w$ is a starvation and/or  
                an unbounded waiting
            }{
                Check $\rho_w \leftarrow \mbox{EF }  \mbox{EG } (w \land (\mbox{EF } \lnot w))$\;
                \eIf{ $\rho_w = \mbox{false}$}{
                    $w$ is a deadlock location
                }{
                    $w$ is a local deadlock,  
                    a starvation and/or  
                    an unbounded waiting location
                }

            }
            
        }
 
    }
}
\caption{\label{algo1}Determining the status of a wait location} 
\end{algorithm}

Thus, in order to detect local deadlocks and starvation (or unbounded waitings) 
in components in \mrl{} specifications we propose the procedure described in 
Algorithm~\ref{algo1}, using the \prism{} model checker 
on the \prism{} representations $\prismsem(\spec)$ of \mrl{} specifications $\spec$.

\subsection{Experimental results}
We applied this procedure to Examples \ref{ex1} and \ref{ex2}.
In order to automatically translate these examples to the \prism{}
language, we extended our compiler \cite{rataj2013probabilistic}
with the emulation of urgent synchronisations, discussed in
Sec.~\ref{sec:prism_semantics}, and with a library with definitions of \mrl{} components.
The results of model checking of formulas and the status of each wait location $s_i'$ 
are shown in Table \ref{tab:results}, where for each component, 
 $s_i'$ is the location added for $s_i$ in Figure \ref{fig:example1_tast} 
in order to emulate urgent communications (see Figure \ref{urgent.fig}).
For Example \ref{ex2}, as the model-checking times are similar, 
we show only locations for which we obtain a different status w.r.t. Example \ref{ex1}.

We may observe that  Example \ref{ex1} presents several locations concerned with both 
local deadlock and starvation (in different contexts), for which starvation disappears in 
Example \ref{ex2} due to the modified timing constraint on the rendering.

\begin{table*}[t]
\begin{center}
\renewcommand*{\arraystretch}{1.1}
\begin{tabular}{|l|ll|c|@{\,\,}l@{ }r@{\,\,}|@{\,\,}l@{ }r@{\,\,}|@{\,\,}c@{}l@{~}r@{}c@{\,\,}|@{\,\,}c@{\,\,}|}
\hline
&&&&\multicolumn{2}{@{}c@{\,\,}|@{\,\,}}{$\phi_w$}  &
\multicolumn{2}{@{}c@{\,\,}|@{\,\,}}{$\psi_w$ }  &
\multicolumn{4}{@{}c@{\,\,}|@{\,\,}}{$\rho_w$} &  \\
example & comp. & $w$ & static set & result & \(t\) [s] 	& result 	& \(t\) [s] 	&~& result & \(t\) [s] && status of $w$ \\[2pt]
\hline
\hline
\multirow{16}{*}{\begin{tabular}{@{}l}Ex.~\ref{ex1} 
			\end{tabular}} & S1
& \(s'_{2}\) & $\SDU$	& false 	& 167 	&  		&  		&  		&&		&&\\[2pt]  
\cline{2-13}
&\multirow{2}{*}{S2}
& \(s'_{2}\) & $\SDU$	& true 	& 228 	& true 	& 184 	&& \textbf{true} 	& 213 	&&  D and S \\[2pt] 
&& \(s'_{3}\) & $\SDU$	& true 	& 228 	& false 	& 156 	&& true 	& 137 	&&  S \\[2pt]  
\cline{2-13}
&\multirow{2}{*}{S3}
& \(s'_{2}\) &$ \SDU$	& true 	& 226 	& true 	& 231 	&& \textbf{true} 	& 176 	&&  D and S \\[2pt] 
&& \(s'_{3}\) & $\SDU$	& false 	& 139 &  &  &&  &  &&\\[2pt]  
\cline{2-13}
&\multirow{2}{*}{F1}
& \(s'_{0}\) & $\SDU$	& false 	& 123 &  &  &&  &  &&\\[2pt] 
&& \(s'_{2}\) & $\onlyS	$	& false 	& 127 &  &  &&  &  &&\\[2pt] 
\cline{2-13}
&\multirow{2}{*}{F2}
& \(s'_{0}\) & $\SDU$	& false 	& 148 &  &  &&  &  &&\\[2pt] 
&& \(s'_{2}\) & $\SDU$ 	& true 	& 214 	& true 	& 167 	&& \textbf{true} 	& 208 	&&  D and S \\[2pt]  
\cline{2-13}
&\multirow{4}{*}{B}
& \(s'_{0}\) & $\SDU$ 	& false 	& 126 &  &  && & && \\[2pt] 
&& \(s'_{1}\) & $\SDU$	& true 	& 298 	& true 	& 198 	&& false 	& 157 	&& D \\[2pt] 
&& \(s'_{2}\) & $\SDU$	& false 	& 137 &  &  &  &&  &&\\[2pt] 
&& \(s'_{4}\) & $\onlyS$	& \textbf{true}	& 202 	& false 	& 149 	&& \textbf{true} 	& 210 	&& S \\[2pt]  
\cline{2-13}
&\multirow{1}{*}{R}
& \(s'_{0}\) & $\onlyS	$	& false & 146 &  &  && & &&\\[2pt]  
\hline
\hline
\multirow{4}{*}{\begin{tabular}{@{}l}Ex.~\ref{ex2} 
			\end{tabular}} & \multirow{1}{*}{S2}
& \(s'_{2}\) & $\SDU$	& true 	& 181 	& true 	& 172 	&& \textbf{false} 	& 110 	&&  D  \\[2pt] 
\cline{2-13}
&\multirow{1}{*}{S3}
& \(s'_{2}\) &$ \SDU$	& true 	& 208 	& true 	& 181 	&& \textbf{false} 	& 108 	&&  D  \\[2pt] 
\cline{2-13}
&\multirow{1}{*}{F2}
& \(s'_{2}\) & $\SDU$ 	& true 	& 163 	& true 	& 158 	&& \textbf{false} & 104 	&&  D  \\[2pt]  
\cline{2-13}
&\multirow{1}{*}{B}
& \(s'_{4}\) & $\onlyS$	& \textbf{false} 	& 94 	& & && & &&  \\[2pt]  
\hline
\end{tabular}
\renewcommand*{\arraystretch}{1}
\end{center}
\caption{\label{tab:results}
Status (D=deadlock, S=starvation) of wait locations in Examples \ref{ex1} and \ref{ex2} obtained with Algorithm \ref{algo1}.
For Example \ref{ex2}, only properties differing from Example \ref{ex1} are shown, the mismatching ones are in bold.
Model checking times \(t\) arisen for a system with AMD Opteron 6234 2.4Ghz and 64GB RAM.}
\end{table*}

\section{Conclusions and perspectives} 

We provided a method allowing to automatically detect 
indefinite waitings in \mrl{} specifications,
and to characterise them as local deadlocks, unbounded waitings or starvation problems, or combinations of them.
We succeeded thanks to a suitable translation of \mrl{} specifications to \prism,
which enabled to model check complex (nested) CTL formulas.
An auxiliary but quite general theoretical contribution of the paper is an efficient (and almost transparent) 
way of expressing urgent communications in \prism{}, which was crucial for our first objective.

The translation from \mrl{} to \tast{}s, \uppaal{} and \prism{} has been automated. 
The \mrl{} compiler can now produce models tuned to specific capabilities
of the target model checker. Yet, while we also support \prism~now, we do not
take advantage of its major feature of checking models which are stochastic.  
Obviously, probability may allow for much more realistic
models and queries within the scope of \mrl{}. We plan thus to introduce unreliable
and stochastic components, to be checked using formalisms like PCTL, i.e., probabilistic CTL.

The computation times of the example models turned out to be quite reasonable, 
with the time constants carefully chosen in order to have a large gcd,
but we should now consider more complex systems, both in terms of structure, 
in terms of interval bound characteristics, 
and in terms of a mixture of stochastic and non-deterministic components.
It could also be considered to introduce several traits of actual programming languages, 
like variables or conditional jumps. In order to stay within the capabilities of model
checkers, though, we might \eg divide a real computer application into functional
blocks, each having, beside an actual implementation, a simplified
specification for \mrl{}, which could be used to \eg checking
properties similar to these discussed here, and consequently discover  
and analyse possible undesired behaviours of the original application.

\paragraph{Acknowledgment}
This work has been partly supported by French ANR project {\sc Synbiotic} and Polish-French project {\sc Polonium}.

\bibliographystyle{eptcs}
\bibliography{eptcs} 

\begin{thebibliography}{10}
\providecommand{\bibitemdeclare}[2]{}
\providecommand{\surnamestart}{}
\providecommand{\surnameend}{}
\providecommand{\urlprefix}{Available at }
\providecommand{\url}[1]{\texttt{#1}}
\providecommand{\href}[2]{\texttt{#2}}
\providecommand{\urlalt}[2]{\href{#1}{#2}}
\providecommand{\doi}[1]{doi:\urlalt{http://dx.doi.org/#1}{#1}}
\providecommand{\bibinfo}[2]{#2}

\bibitemdeclare{inproceedings}{alur:90}
\bibitem{alur:90}
\bibinfo{author}{Rajeev \surnamestart Alur\surnameend} \&
  \bibinfo{author}{David~L. \surnamestart Dill\surnameend}
  (\bibinfo{year}{1990}): \emph{\bibinfo{title}{Automata for modeling real-time
  systems}}.
\newblock In: {\sl \bibinfo{booktitle}{International Colloquium on Algorithms,
  Languages, and Programming (ICALP) 1990}}, {\sl \bibinfo{series}{LNCS}}
  \bibinfo{volume}{443}, \bibinfo{publisher}{Springer}, pp.
  \bibinfo{pages}{322--335}, \doi{10.1007/BFb0032042}.

\bibitemdeclare{inproceedings}{alur:91}
\bibitem{alur:91}
\bibinfo{author}{Rajeev \surnamestart Alur\surnameend} \&
  \bibinfo{author}{David~L. \surnamestart Dill\surnameend}
  (\bibinfo{year}{1991}): \emph{\bibinfo{title}{The theory of timed automata}}.
\newblock In: {\sl \bibinfo{booktitle}{Real Time: Theory in Practice (REX
  Workshop)}}, {\sl \bibinfo{series}{LNCS}} \bibinfo{volume}{600},
  \bibinfo{publisher}{Springer}, pp. \bibinfo{pages}{45--73},
  \doi{10.1007/BFb0031987}.

\bibitemdeclare{article}{alur:94}
\bibitem{alur:94}
\bibinfo{author}{Rajeev \surnamestart Alur\surnameend} \&
  \bibinfo{author}{David~L. \surnamestart Dill\surnameend}
  (\bibinfo{year}{1994}): \emph{\bibinfo{title}{A theory of timed automata}}.
\newblock {\sl \bibinfo{journal}{Theoretical Computer Science}}
  \bibinfo{volume}{126}(\bibinfo{number}{2}), pp. \bibinfo{pages}{183--235},
  \doi{10.1016/0304-3975(94)90010-8}.

\bibitemdeclare{techreport}{arcile2014}
\bibitem{arcile2014}
\bibinfo{author}{Johan \surnamestart Arcile\surnameend} (\bibinfo{year}{2014}):
  \emph{\bibinfo{title}{Impl\'ementation d'un outil de compilation des
  sp\'ecifications MIRELA vers les automates temporis\'es au format UPPAAL
  (XML)}}.
\newblock \bibinfo{type}{Rapport de stage L3},
  \bibinfo{institution}{D\'epartement Informatique, Universit\'e d'Evry,
  France}.

\bibitemdeclare{inproceedings}{bauer01}
\bibitem{bauer01}
\bibinfo{author}{Martin \surnamestart Bauer\surnameend}, \bibinfo{author}{Bernd
  \surnamestart Bruegge\surnameend}, \bibinfo{author}{Gudrun \surnamestart
  Klinker\surnameend}, \bibinfo{author}{Asa \surnamestart
  MacWilliams\surnameend}, \bibinfo{author}{Thomas \surnamestart
  Reicher\surnameend}, \bibinfo{author}{Stephan \surnamestart Riss\surnameend},
  \bibinfo{author}{Christian \surnamestart Sandor\surnameend} \&
  \bibinfo{author}{Martin \surnamestart Wagner\surnameend}
  (\bibinfo{year}{2001}): \emph{\bibinfo{title}{Design of a Component-Based
  Augmented Reality Framework}}.
\newblock In: {\sl \bibinfo{booktitle}{Proceedings of the International
  Symposium on Augmented Reality (ISAR)}}, \doi{10.1109/ISAR.2001.970514}.

\bibitemdeclare{inproceedings}{JY}
\bibitem{JY}
\bibinfo{author}{Mehdi \surnamestart Chouiten\surnameend},
  \bibinfo{author}{Christophe \surnamestart Domingues\surnameend},
  \bibinfo{author}{Jean-Yves \surnamestart Didier\surnameend},
  \bibinfo{author}{Samir \surnamestart Otmane\surnameend} \&
  \bibinfo{author}{Malik \surnamestart Mallem\surnameend}
  (\bibinfo{year}{2012}): \emph{\bibinfo{title}{Distributed mixed reality for
  remote underwater telerobotics exploration}}.
\newblock In: {\sl \bibinfo{booktitle}{Virtual Reality International
  Conference, VRIC '12}}, \bibinfo{publisher}{ACM}, \bibinfo{address}{France},
  pp. \bibinfo{pages}{1:1--1:6}, \doi{10.1145/2331714.2331716}.

\bibitemdeclare{inproceedings}{sand13}
\bibitem{sand13}
\bibinfo{author}{Raymond \surnamestart Devillers\surnameend},
  \bibinfo{author}{Jean-Yves \surnamestart Didier\surnameend} \&
  \bibinfo{author}{Hanna \surnamestart Klaudel\surnameend}
  (\bibinfo{year}{2013}): \emph{\bibinfo{title}{Implementing Timed Automata
  Specifications: The" Sandwich" Approach}}.
\newblock In: {\sl \bibinfo{booktitle}{13th International Conference on
  Application of Concurrency to System Design (ACSD), 2013}},
  \bibinfo{organization}{IEEE}, pp. \bibinfo{pages}{226--235},
  \doi{10.1109/ACSD.2013.26}.

\bibitemdeclare{inproceedings}{issre14}
\bibitem{issre14}
\bibinfo{author}{Raymond \surnamestart Devillers\surnameend},
  \bibinfo{author}{Jean{-}Yves \surnamestart Didier\surnameend},
  \bibinfo{author}{Hanna \surnamestart Klaudel\surnameend} \&
  \bibinfo{author}{Johan \surnamestart Arcile\surnameend}
  (\bibinfo{year}{2014}): \emph{\bibinfo{title}{Deadlock and Temporal
  Properties Analysis in Mixed Reality Applications}}.
\newblock In: {\sl \bibinfo{booktitle}{25th {IEEE} International Symposium on
  Software Reliability Engineering, {ISSRE} 2014, Naples, Italy, November 3-6,
  2014}}, \bibinfo{publisher}{{IEEE}}, pp. \bibinfo{pages}{55--65},
  \doi{10.1109/ISSRE.2014.33}.

\bibitemdeclare{article}{mir08}
\bibitem{mir08}
\bibinfo{author}{Jean-Yves \surnamestart Didier\surnameend},
  \bibinfo{author}{Bachir \surnamestart Djafri\surnameend} \&
  \bibinfo{author}{Hanna \surnamestart Klaudel\surnameend}
  (\bibinfo{year}{2009}): \emph{\bibinfo{title}{The MIRELA framework: modeling
  and analyzing mixed reality applications using timed automata}}.
\newblock {\sl \bibinfo{journal}{Journal of Virtual Reality and Broadcasting}}
  \bibinfo{volume}{6}(\bibinfo{number}{1}).

\bibitemdeclare{inproceedings}{mir13}
\bibitem{mir13}
\bibinfo{author}{Jean-Yves \surnamestart Didier\surnameend},
  \bibinfo{author}{Hanna \surnamestart Klaudel\surnameend},
  \bibinfo{author}{Mathieu \surnamestart Moine\surnameend} \&
  \bibinfo{author}{Raymond \surnamestart Devillers\surnameend}
  (\bibinfo{year}{2013}): \emph{\bibinfo{title}{An improved approach to build
  safer mixed reality systems by analysing time constraints}}.
\newblock In: {\sl \bibinfo{booktitle}{Proceedings of the 5th Joint Virtual
  Reality Conference}}.

\bibitemdeclare{article}{endres05}
\bibitem{endres05}
\bibinfo{author}{Christoph \surnamestart Endres\surnameend},
  \bibinfo{author}{Andreas \surnamestart Butz\surnameend} \&
  \bibinfo{author}{Asa \surnamestart MacWilliams\surnameend}
  (\bibinfo{year}{2005}): \emph{\bibinfo{title}{A Survey of Software
  Infrastructures and Frameworks for Ubiquitous Computing}}.
\newblock {\sl \bibinfo{journal}{Mobile Information Systems Journal}}
  \bibinfo{volume}{1}(\bibinfo{number}{1}), pp. \bibinfo{pages}{41--80}.

\bibitemdeclare{article}{figueroa2008}
\bibitem{figueroa2008}
\bibinfo{author}{Pablo \surnamestart Figueroa\surnameend},
  \bibinfo{author}{Walter~F \surnamestart Bischof\surnameend},
  \bibinfo{author}{Pierre \surnamestart Boulanger\surnameend},
  \bibinfo{author}{H~James \surnamestart Hoover\surnameend} \&
  \bibinfo{author}{Robyn \surnamestart Taylor\surnameend}
  (\bibinfo{year}{2008}): \emph{\bibinfo{title}{Intml: A dataflow oriented
  development system for virtual reality applications}}.
\newblock {\sl \bibinfo{journal}{Presence: Teleoperators and Virtual
  Environments}} \bibinfo{volume}{17}(\bibinfo{number}{5}), pp.
  \bibinfo{pages}{492--511}, \doi{10.1162/pres.17.5.492}.

\bibitemdeclare{article}{figueroa2004}
\bibitem{figueroa2004}
\bibinfo{author}{Pablo \surnamestart Figueroa\surnameend},
  \bibinfo{author}{J~\surnamestart Hoover\surnameend} \&
  \bibinfo{author}{Pierre \surnamestart Boulanger\surnameend}
  (\bibinfo{year}{2004}): \emph{\bibinfo{title}{Intml concepts}}.
\newblock {\sl \bibinfo{journal}{University of Alberta. Computing Science
  Department, Tech. Rep}}.

\bibitemdeclare{techreport}{haller03}
\bibitem{haller03}
\bibinfo{author}{Michael \surnamestart Haller\surnameend},
  \bibinfo{author}{J\"{u}rgen \surnamestart Zauner\surnameend},
  \bibinfo{author}{Werner \surnamestart Hartmann\surnameend} \&
  \bibinfo{author}{Thomas \surnamestart Luckeneder\surnameend}
  (\bibinfo{year}{2003}): \emph{\bibinfo{title}{A generic framework for a
  training application based on Mixed Reality.}}
\newblock \bibinfo{type}{Technical Report}, \bibinfo{institution}{Upper Austria
  University of Applied Sciences}, \bibinfo{address}{Hagenberg, Austria}.

\bibitemdeclare{article}{hughes2005mixed}
\bibitem{hughes2005mixed}
\bibinfo{author}{Charles~E \surnamestart Hughes\surnameend},
  \bibinfo{author}{Christopher~B \surnamestart Stapleton\surnameend},
  \bibinfo{author}{Darin~E \surnamestart Hughes\surnameend} \&
  \bibinfo{author}{Eileen~M \surnamestart Smith\surnameend}
  (\bibinfo{year}{2005}): \emph{\bibinfo{title}{Mixed reality in education,
  entertainment, and training}}.
\newblock {\sl \bibinfo{journal}{Computer Graphics and Applications, IEEE}}
  \bibinfo{volume}{25}(\bibinfo{number}{6}), pp. \bibinfo{pages}{24--30},
  \doi{10.1109/MCG.2005.139}.

\bibitemdeclare{article}{kwiatkowska2004probabilistic}
\bibitem{kwiatkowska2004probabilistic}
\bibinfo{author}{M.~\surnamestart Kwiatkowska\surnameend},
  \bibinfo{author}{G.~\surnamestart Norman\surnameend} \&
  \bibinfo{author}{D.~\surnamestart Parker\surnameend} (\bibinfo{year}{2004}):
  \emph{\bibinfo{title}{Probabilistic Symbolic Model Checking with {PRISM}: A
  Hybrid Approach}}.
\newblock {\sl \bibinfo{journal}{International Journal on Software Tools for
  Technology Transfer (STTT)}} \bibinfo{volume}{6}(\bibinfo{number}{2}), pp.
  \bibinfo{pages}{128--142}, \doi{10.1007/s10009-004-0140-2}.

\bibitemdeclare{inproceedings}{latoschik2002}
\bibitem{latoschik2002}
\bibinfo{author}{Marc~Erich \surnamestart Latoschik\surnameend}
  (\bibinfo{year}{2002}): \emph{\bibinfo{title}{Designing transition networks
  for multimodal VR-interactions using a markup language}}.
\newblock In: {\sl \bibinfo{booktitle}{Proceedings of the 4th IEEE
  International Conference on Multimodal Interfaces}},
  \bibinfo{organization}{IEEE Computer Society}, p. \bibinfo{pages}{411},
  \doi{10.1109/ICMI.2002.1167030}.

\bibitemdeclare{incollection}{navarre2005}
\bibitem{navarre2005}
\bibinfo{author}{David \surnamestart Navarre\surnameend},
  \bibinfo{author}{Philippe \surnamestart Palanque\surnameend},
  \bibinfo{author}{R{\'e}mi \surnamestart Bastide\surnameend},
  \bibinfo{author}{Amelie \surnamestart Schyn\surnameend},
  \bibinfo{author}{Marco \surnamestart Winckler\surnameend},
  \bibinfo{author}{Luciana~P \surnamestart Nedel\surnameend} \&
  \bibinfo{author}{Carla~MDS \surnamestart Freitas\surnameend}
  (\bibinfo{year}{2005}): \emph{\bibinfo{title}{A formal description of
  multimodal interaction techniques for immersive virtual reality
  applications}}.
\newblock In: {\sl \bibinfo{booktitle}{Human-Computer Interaction-INTERACT
  2005}}, \bibinfo{publisher}{Springer}, pp. \bibinfo{pages}{170--183},
  \doi{10.1007/11555261\_17}.

\bibitemdeclare{inproceedings}{piekarski03}
\bibitem{piekarski03}
\bibinfo{author}{Wayne \surnamestart Piekarski\surnameend} \&
  \bibinfo{author}{Bruce~H. \surnamestart Thomas\surnameend}
  (\bibinfo{year}{2003}): \emph{\bibinfo{title}{An Object-Oriented Software
  Architecture for {3D} Mixed Reality Applications}}.
\newblock In: {\sl \bibinfo{booktitle}{ISMAR '03: Proceedings of the The 2nd
  IEEE and ACM International Symposium on Mixed and Augmented Reality}},
  \bibinfo{publisher}{IEEE Computer Society}, \bibinfo{address}{Washington, DC,
  USA}, p. \bibinfo{pages}{247}, \doi{10.1109/ISMAR.2003.1240708}.

\bibitemdeclare{article}{rataj2013probabilistic}
\bibitem{rataj2013probabilistic}
\bibinfo{author}{Artur \surnamestart Rataj\surnameend} (\bibinfo{year}{2013}):
  \emph{\bibinfo{title}{Translation of probabilistic games in {J2TADD}}}.
\newblock {\sl \bibinfo{journal}{Theoretical and Applied Informatics}}
  \bibinfo{volume}{25}(\bibinfo{number}{3/4}).

\bibitemdeclare{inproceedings}{reitmayr01}
\bibitem{reitmayr01}
\bibinfo{author}{Gerhard \surnamestart Reitmayr\surnameend} \&
  \bibinfo{author}{Dieter \surnamestart Schmalstieg\surnameend}
  (\bibinfo{year}{2001}): \emph{\bibinfo{title}{An open software architecture
  for virtual reality interaction}}.
\newblock In: {\sl \bibinfo{booktitle}{Proceedings of the ACM symposium on
  Virtual reality software and technology}}, \bibinfo{publisher}{ACM Press},
  pp. \bibinfo{pages}{47--54}, \doi{10.1145/505008.505018}.

\bibitemdeclare{inproceedings}{sandor2001}
\bibitem{sandor2001}
\bibinfo{author}{Christian \surnamestart Sandor\surnameend},
  \bibinfo{author}{Thomas \surnamestart Reicher\surnameend} et~al.
  (\bibinfo{year}{2001}): \emph{\bibinfo{title}{CUIML: A Language for the
  Generation of Multimodal Human-Computer Interfaces}}.
\newblock In: {\sl \bibinfo{booktitle}{Proceedings of the European UIML
  conference}}, \bibinfo{volume}{124}.

\bibitemdeclare{misc}{uppaal4.1}
\bibitem{uppaal4.1}
\emph{\bibinfo{title}{UPPAAL}}.
\newblock \bibinfo{howpublished}{\url{http://www.uppaal.org/}}.

\bibitemdeclare{article}{visser2003modelchecking}
\bibitem{visser2003modelchecking}
\bibinfo{author}{Willem \surnamestart Visser\surnameend},
  \bibinfo{author}{Klaus \surnamestart Havelund\surnameend},
  \bibinfo{author}{Guillaume \surnamestart Brat\surnameend},
  \bibinfo{author}{Seungjoon \surnamestart Park\surnameend} \&
  \bibinfo{author}{Flavio \surnamestart Lerda\surnameend}
  (\bibinfo{year}{2003}): \emph{\bibinfo{title}{Model Checking Programs}}.
\newblock {\sl \bibinfo{journal}{Automated Software Engineering Journal}}
  \bibinfo{volume}{10}(\bibinfo{number}{2}), \doi{10.1023/A:1022920129859}.

\bibitemdeclare{techreport}{Waez2011}
\bibitem{Waez2011}
\bibinfo{author}{Md~Tawhid~Bin \surnamestart Waez\surnameend},
  \bibinfo{author}{J\"urgen \surnamestart Dingel\surnameend} \&
  \bibinfo{author}{Karen \surnamestart Rudie\surnameend}
  (\bibinfo{year}{2011}): \emph{\bibinfo{title}{Timed Automata for the
  Development of Real-Time Systems}}.
\newblock \bibinfo{type}{Research Report} \bibinfo{number}{2011-579},
  \bibinfo{institution}{Queen's University -- School of Computing, Canada}.

\end{thebibliography}

\end{document}